\documentclass[10pt]{iopart}
\begin{document}

\title{Landauer Formula without Landauer's Assumptions}

\author{Mukunda P. Das\footnote[3]
{To whom correspondence should be addressed
(mukunda.das@anu.edu.au)}
and Frederick Green}

\address{Department of Theoretical Physics,
Institute of Advanced Studies,\\
The Australian National University,
Canberra, ACT 0200, Australia}

\begin{abstract}
The Landauer formula for dissipationless conductance lies
at the heart of modern electronic transport, yet it
remains without a clear microscopic basis.
We analyze the Landauer formula microscopically, and give
a straightforward quantum kinetic derivation for open systems.
Some important experimental implications follow.
These lie beyond the Landauer result as popularly received.
\end{abstract}






In 1957 Rolf Landauer published a prescient
interpretation of metallic resistivity
\cite{ldr57}.
It heralded one of the most dramatic
predictions of modern
condensed-matter physics: the perfect quantization,
in steps of $2 e^2/h$, of
electrical conductance in one-dimensional
metallic channels
\cite{imry}.
Such quantization is quite independent
of the material properties of the contact and
of its leads. It is universal insofar as one may
validly neglect the disruptive influences of
{\em inelastic dissipation} within the transport process.

Landauer argued that the current, not the applied
electromotive voltage, should be understood as the
active probe by which a device reveals its conductance.
The observed carrier flux is understood as a
kind of diffusive flow, tending to shift carriers
from a ``high''- to a ``low''-density reservoir (lead).
In the mesoscopic realm, this flow between
leads is conditioned by the intervening device channel,
which presents a quantum tunnelling barrier to
the non-interacting electrons making up the flux.

The Landauer formula, then, has two cardinal tenets:
\smallskip

(i) current is the flow of independent and degenerate
electrons as they follow a nominal density gradient
across reservoirs, and
\smallskip

(ii) conductance is {\em lossless} transmission through an
interposed quantum barrier.
\smallskip

\noindent
These underpin Landauer's
assumptions, namely that
\smallskip

(a) transport ensues when a pair
of leads connected to the device are set to different
chemical potentials $\mu_{\rm L}, \mu_{\rm R}$;
\smallskip

(b) the density mismatch due to $\mu_{\rm L} - \mu_{\rm R}$
sustains the current;
\smallskip

(c) $\mu_{\rm L} - \mu_{\rm R}$
is the applied voltage across the device;
\smallskip

(d) the Fermi energy
is much larger than the thermal and electrical energies; and
\smallskip

(e) there are no inelastic processes to dissipate the
electrical energy gained by the electrons.
\smallskip

Energy dissipation does not appear in the
classic Landauer derivation
\cite{ldr57,rmpIL}.
For a sample of mesoscopic dimensions,
the model admits only elastic barrier scattering
and excludes any role for inelastic processes within
the active device and its interfaces.
Yet it is dissipative inelastic scattering, and
that alone, which ensures the energetic stability
of resistive transport, and hence a steady state for conduction.

Finite conductance and electrical energy loss are
indivisible phenomena.
The fundamental expression of their basic
inseparability is the fluctuation-dissipation relation
\cite{nvk}.
This establishes the equivalence of the mean-square fluctuation
strength for the current and the conductance coefficient $G$
in the the energy dissipation rate $P = GV^2$, where $V$ is
the applied voltage.

There is a missing link between Landauer's
universal -- and lossless -- conductance formula,
which has been critical in the development of mesoscopic science
\cite{rmpIL},
and the dissipative inelastic processes that are absolutely vital to
the microscopic origin of resistance.
Repeated attempts have been made to obtain
the Landauer formula from microscopic-like arguments
\cite{imry,fisherlee,barstone};
see also Ciraci et al
\cite{ciraci}.
However, a convincing resolution has not yet materialized
\cite{sols,kk}.
The absence of so crucial a connection is a
puzzling theoretical conundrum for Landauer's
approach to mesoscopics, which is otherwise
so empirically compelling.

In this letter we answer the question:
How can the Landauer formula, in seemingly bypassing
{\em all} inelastic processes, predict a
finite -- invariably dissipative -- conductance
that fulfils the fluctuation-dissipation theorem (FDT)?
Below we offer a straightforward microscopic interpretation
of Landauer's result, for
a mesoscopic contact open to the macroscopic environment.

Our treatment differs from all earlier attempts by directly
addressing the essential physics of dissipation.
To obtain conductance quantization within an open
contact, the explicit interplay of elastic
and dissipative processes is necessary {\em and sufficient}.
Neglect of either mechanism, in favour of the other,
negates the formula's microscopic basis. Both kinds of
scattering are needed.

We also show that the traditional Landauer assumptions of
pseudo-diffusive current and lossless scattering are not
required in a first-principles analysis of Landauer conductance.
Our model relies solely upon orthodox quantum kinetics,
as embodied in the microscopic Kubo-Greenwood (KG) formalism
\cite{kubo,zimod}.
The KG formulation automatically guarantees the FDT;
it is not invoked as an additional hypothesis.
Both dissipative and lossless scattering appear within the
resulting fluctuation-dissipation relation,
and both are assigned {\em equal} physical importance.

First, we briefly recall the KG formula
and the essential charge conservation built into it.
Next we discuss the form of the KG relaxation time, which
fixes the conductance. Finally, we show how the physical
constraints on a one-dimensional {\em open} ballistic channel,
connected to macroscopic leads, leads naturally to
Landauer's ideal quantized conductance. We go on to examine
some of the measurable effects of device non-ideality
on the Landauer conductance.

The Kubo-Greenwood theory
\cite{kubo,zimod}
decribes the carriers' full many-body density matrix. All of the
transport and fluctuation properties are contained within it.
Thus, the conductivity for the system appears as the trace of
the current-current correlation function:

\begin{equation}
\sigma(t) = {ne^2\over m^*}\int^t_0 {\cal C}_{vv}(t) dt.
\label{eq1}
\end{equation}

\noindent
Here $n$ is the carrier density and $m^*$ the effective mass.
The velocity auto-correlation has the canonical form
\begin{equation}
C_{vv}(t) = {{\langle [v(t),v(0)] \rangle}\over
             {\langle v(0)^2\rangle}} 
\sim \exp(-t/\tau_m)
\label{eq2}
\end{equation}

\noindent
where the expectations trace over the equilibrium density matrix
(this gives the leading, linear term in the expansion of the
non-equilibrium response).
For long times, the characteristic relaxation rate
$1/\tau_m$ yields the dominant decay of the correlation.

The asymptotic relaxation rate subsumes, {\em on an equal footing},
the microscopic contribution from every physically relevant
collision mechanism.
Moving now to the long-time form of Equation (\ref{eq1}),
the conductivity becomes

\begin{equation}
\sigma \to {ne^2\tau_m\over m^*}.
\label{eq3}
\end{equation}

\noindent
This is the celebrated Drude formula.

Equation (\ref{eq1}) embodies the fluctuation-dissipation
relation. In addition, its KG structure ensures that
charge conservation is rigidly satisfied in the large,
as well as locally
\cite{sols}. 
This is an absolute prerequisite for open conductors
as they exchange carriers freely with the outside. 

Since Landauer's classic result applies to transport in a
one-dimensional metallic wire,
we examine Eq. (\ref{eq3}) in one dimension (1D), for
a single metallic sub-band (channel) within the wire.
In the degenerate limit the density is $n = 2k_{\rm F}/\pi$
in terms of the Fermi wave-number $k_{\rm F}$.
The conductance over a sample of length $L$ becomes

\begin{equation}
G \equiv {\sigma\over L} = {2k_{\rm F}e^2\over \pi L m^*} \tau_m
= {2e^2\over h}{\left( {2\hbar k_{\rm F}\over L m^*} \tau_m \right)}
\equiv {2e^2\over h}{\cal T}_{\rm KG},
\label{eq4}
\end{equation}

\noindent
in which the transmission coefficient
${\cal T}_{\rm KG}= 2v_{\rm F}\tau_m/L$
is proportional to the ratio of the overall scattering
length, $v_{\rm F}\tau_m$, to the operational length of the
system.

Crucially, the many-body collisions
(phonon emission, Coulomb scattering, etc.)
that redistribute the carriers' energy gain 
and cause dissipation are incorporated in $\tau_m$ alongside
elastic impurity and barrier scattering.
While elastic effects are explicitly invoked by the Landauer model,
dissipative ones are neglected. It is dissipation that
stabilizes the transport and substantiates the
fluctuation-dissipation relation, Eq. (\ref{eq1}).

Equation (\ref{eq4}) is fully consistent with
the Landauer formula, which is identical to it except that,
in the accepted treatment, its transmission parameter
${\cal T}$ is ideal: ${\cal T} = 1$.
In cases where ${\cal T}$ is {\em not} ideal the Landauer picture
assumes that the non-ideality is due solely to elastic back-scattering
from the barrier, but does not facilitate the actual
computation of ${\cal T}$. When inelastic scattering dominates,
this picture is inapplicable
\cite{agrait}.

Let us take a simple model for  ${\cal T}_{\rm KG}$.
The wire is ballistic (impurity-free), and it is uniform;
by Poisson's equation, so are the driving field and
carrier distribution set up within it.
At distance $L$ apart lie the wire-lead interfaces where the current
is, in effect, injected and extracted by an outside generator.
We observe that $L$ is {\em not} a lithographically precise
dimension. It characterizes the maximum physical scale
for any collision process to occur in the entire mesoscopic
assembly (the open wire, the interfaces, and the reservoirs
are one whole system).
Note also that it is the external supply and removal
of the current that explicitly energizes the open system
\cite{sols}.
There is no appeal in Eq. (\ref{eq1})
to chemical-potential differences in any way, shape, or form.

The wire-reservoir interfaces are zones of strong
elastic scattering with impurities in the leads
(the relaxation time is $\tau_{\rm el}$); equally they are sites for 
strong {\it dissipative} interactions with the background modes
excited by the influx and efflux of carriers from the current source
(the relaxation time is $\tau_{\rm in}$).
The scattering mechanisms are stochastically independent, so that
Matthiessen's rule applies:

\begin{equation}
{1\over \tau_m} = {1\over \tau_{\rm el}} +  {1\over \tau_{\rm in}}.
\label{eq5}
\end{equation}

\noindent
The mean free path (MFP) associated with the elastic collisions
is obviously $L$, since by hypothesis that is the operational
size of our impurity-free wire. Therefore $\tau_{\rm el} = L/v_{\rm F}$
for carriers at the Fermi level.
By the same token, the MFP for inelastic scattering cannot be
{\it greater} than $L$, though it may well be less at high
currents\footnote
{In a ``diffusive'' wire, containing many elastic scattering centres,
the complementary scenario holds: the wire length $L$ represents 
a maximum scale for inelastic scattering, so that
$\tau_{\rm el} < \tau_{\rm in} \leq L/v_{\rm F}.$}.
Then

\begin{equation}
\tau_{\rm in} \leq \tau_{\rm el} = L/v_{\rm F}.
\label{eq6}
\end{equation}

\noindent
We conclude that

\begin{equation}
{\cal T}_{\rm KG}
= {2v_{\rm F}\over L}{\left( {{\tau_{\rm el} \tau_{\rm in}}\over
            {\tau_{\rm el} + \tau_{\rm in}}} \right)}
= {2\tau_{\rm in}\over {\tau_{\rm in} + L/v_{\rm F}}}
\leq 1.
\label{eq7}
\end{equation}

It is the direct competition between the elastic processes
in the mesoscopic system (as a ballistic structure, its
elastic mean free path is also its characteristic length)
and the dissipative processes (ideally restricted to the
current injection/extraction areas bounding $L$,
but also liable to intrude into the interior)
that determines the physical, and measurable, transmission
through the sample.

What is the optimum outcome for Eq. (\ref{eq7}), and what
does it yield for the conductance? The maximum value of
${\cal T}_{\rm KG}$ is unity, and it is attained precisely when

\begin{equation}
\tau_{\rm in} = \tau_{\rm el} = L/v_{\rm F}.
\label{eq8}
\end{equation}

\noindent
In other words, no inelastic events intrude into the
core of the wire; they all occur at the interfaces.
From Eq. (\ref{eq4}) one easily discerns the corresponding
value of $G$ for this open, maximally ballistic 1D wire.
It is nothing but the Landauer conductance $G_0 = 2e^2/h$.

This establishes our key result.
As with Landauer's derivation, we base
it on two hypotheses: (i) that the wire is uniform,
and (ii) that its 1D conduction sub-bands are well enough
separated in energy that each can be treated independently.

Our account of the Landauer formula makes no use at all
of the three other assumptions that are traditionally relied
upon to establish the formula. They are:

\begin{itemize}
\item
{\it That a mesoscopic current flows only when there is
a density mismatch between carrier reservoirs, held at
different chemical potentials}.

\item
{\it That coherent elastic scattering is the exclusive
transmission mechanism mediating the conductance}.

\item
{\it That dissipation in an open conductor
(accepted as vital in order to save the FDT)
is a remote effect deep in the reservoirs,
of no physical consequence for transport}.
\end{itemize}

\noindent
We have demonstrated that these assumptions are superfluous
in obtaining Landauer's result. A fourth key assumption remains:

\begin{itemize}
\item
{\it That the quantized-conductance formula requires linear
response in a degenerate channel}.
\end{itemize}

\noindent
We now show that this hypothesis too is not required for
understanding the microscopic basis of mesoscopic conductance.

A standard kinetic approach suffices to describe the
carriers in a ballistic and uniform 1D conductor
\cite{fnl}.
In steady state, with a driving field $E$ (to be determined),
our model carrier distribution function $f_k$
in wave-vector space $\{{k\}}$ obeys the transport equation

\begin{equation}
{eE\over \hbar}
{\partial f_k\over \partial k}
=
-{1\over \tau_{\rm in}(\varepsilon_k)}
{\left( f_k -
{{\langle \tau_{\rm in}^{-1} f \rangle}\over 
 {\langle \tau_{\rm in}^{-1} f^{\rm eq} \rangle}}
f^{\rm eq}_k
\right)}
-{1\over \tau_{\rm el}(\varepsilon_k)}
{ {f_k - f_{-k}}\over 2 }.
\label{eq9}
\end{equation}

\noindent
The scattering times $\tau_{\rm in}(\varepsilon_k)$
and $\tau_{\rm el}(\varepsilon_k)$
are in general dependent on the band energy $\varepsilon_k$.

The properties of Eq. (\ref{eq9}) impact
directly upon the measurable transport behaviour.
First, one and only one chemical potential $\mu$
enters the problem, via the equilibrium Fermi-Dirac
distribution at temperature $T$:

\[
f^{\rm eq}_k = 1/{\{ 1 +
\exp[(\varepsilon_k + \varepsilon_i - \mu)/k_{\rm B}T] \}}.
\]

\noindent
Quite generally, this is the reference state for computing
the non-equilibrium function $f_k$
\cite{gd2}
(here $\varepsilon_i$ is the energy threshold of the sub-band).
The applicability of Eq. (\ref{eq9}) stretches over
the entire range of density $n = {\langle f \rangle}$,
from classical to strongly degenerate.

Second, the kinetic equation is microscopically conserving.
On the right-hand side of Eq. (\ref{eq9}),
the leading, inelastic, collision term has a restoring contribution
proportional to the expectation

\[
{\langle \tau_{\rm in}^{-1} f \rangle}
= \int^{\infty}_{-\infty}
{2dk\over 2\pi} \tau_{\rm in}^{-1}(\varepsilon_k) f_k.
\]

\noindent
Finally, the second term on the right of Eq. (\ref{eq9})
represents the elastic collisions, acting to restore
symmetry to $f_k$.
Both the elastic and inelastic terms satisfy gauge invariance.

The transport equation is analytically solvable when
the collision times are independent of the electronic band energy
\cite{fnl}.
At low currents, the solution has a transport behaviour
identical to the Kubo-Greenwood formula described above.
At high currents, for which neither the KG nor the Landauer
expressions strictly apply, the kinetic solution remains
tractable.

We now obtain $G$. As we have recalled, the common derivation
of the Landauer conductance posits a highly degenerate
electronic sub-band
\cite{rmpIL}.
That is, we are in the zero-temperature limit.
If the sub-band is populated even vestigially, the
ideal conductance $G = G_0$ always emerges;
but if the band is empty (the only other possibility
at zero temperature), there is no transport and $G = 0$.
There is no room for the intermediate band-threshold
state that is expected at finite temperature. 

The above approach cannot be used in a realistic setting,
where the Fermi energy may well match the thermal energy.
Experimentally
{\cite{vanw,depic},
the carrier density in a 1D channel is controlled via
an adjacent gate. As the gate-bias voltage becomes more positive,
the electron population undergoes a continuous change,
from a low-density classical regime to
a high-density degenerate one.

This classical-to-quantum
transition is readily accommodated.
Classically, the elastic mean free path no longer
scales with the Fermi velocity, but with the thermal velocity
$v_{\rm th} = \sqrt{ 2k_{\rm B}T/m^*}$.
In the general case, $\tau_{\rm el}$
is given by the expression

\begin{equation}
\tau_{\rm el}(n, T) = {L\over {\overline v(n, T)}}
\equiv L{n\over {\langle |v| f^{\rm eq} \rangle}}.
\label{eq10}
\end{equation}

\noindent
For a sparse, classical channel population,
the characteristic mean velocity ${\overline v(n, T)}$ goes to
$v_{\rm th}$.
For a dense and thus degenerate population,
${\overline v} = v_{\rm F} = \sqrt{2(\mu - \varepsilon_i)/m^*}$,
which holds for Eq. (\ref{eq7}) above.
We can then extend Eqs. (\ref{eq4}) and (\ref{eq8})
for $G$ and ${\cal T}_{\rm KG}$ to the whole regime
of densities $n_i$ in the {\em i}th sub-band
accessible at finite temperature:

\begin{equation}
G_i = G_0 {\left( {hn_i\over 2m^*{\overline v(n_i, T)}} \right)}
{\left( 1 - { 1\over {1 + \tau_{\rm in}/\tau_{\rm el}(n_i,T)} } \right)},
\label{eq11}
\end{equation}

\noindent
where $v_{\rm F}$ is replaced with its equivalent
expression in 1D: $v_{\rm F} = \hbar k_{\rm F}/m^* = hn_i/4m^*$.

When the system is at low density ($\mu - \varepsilon_i \ll k_{\rm B}T$)
the conductance vanishes with $n_i$. When the system is degenerate
($\mu - \varepsilon_i \gg k_{\rm B}T$) the conductance reaches
a plateau, which is ideally quantized
when $\tau_{\rm in} =\tau_{\rm el}$. In between, it rises smoothly
as the chemical potential and density are systematically
swept from much below
the sub-band threshold $\varepsilon_i$ to much above it.

\begin{figure}
\input psfig.sty
\centerline{\hskip 20mm\psfig{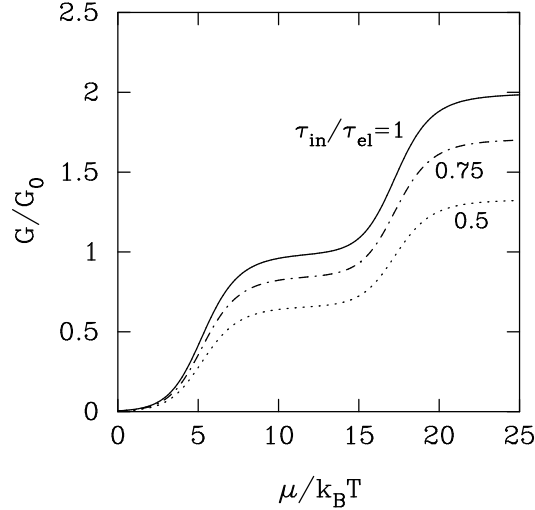}}
\vskip -20mm
\caption{
Conductance of a one-dimensional ballistic wire, computed
with the kinetic model of Eq. (\ref{eq9}).
We show $G$ scaled to the Landauer quantum $G_0$, as a function of
chemical potential $\mu$ in units of thermal energy.
$G$ exhibits strong shoulders as $\mu$ successively
crosses the sub-band-energy thresholds set at
$\varepsilon_1 = 5k_{\rm B}T$ and $\varepsilon_2 = 17k_{\rm B}T$.
Well above each threshold, sub-band electrons are strongly
degenerate and the conductance tends to a well defined
quantized plateau; well below each threshold, the population
and its contribution to $G$ vanish as
$\exp[-(\varepsilon_i-\mu)/k_{\rm B}T]$.
Solid line: $G$ in a ballistic channel. This is the ideal limit
for which the collision-time ratio $\tau_{\rm in}/\tau_{\rm el}$ is unity.
Dot-dashed line: non-ideal case for $\tau_{\rm in}/\tau_{\rm el} = 0.75$.
Note how the increased inelastic scattering brings down the
plateaux. Dotted line: the case of
$\tau_{\rm in}/\tau_{\rm el} = 0.5$. The departure from ideality
is now pronounced.}
\end{figure}

The result is depicted in Figure 1.
We see there the total conductance of a 1D wire,

\[
G = \sum_i G_i{\Bigl( (\mu - \varepsilon_i)/k_{\rm B}T \Bigr)},
\]

\noindent
made up of its individual sub-band contributions
computed from Eq. (\ref{eq11}), with full temperature dependence.
The shoulders at the two sub-band thresholds are clear.
In an idealized scenario (recall Eq. (\ref{eq8})),
the characteristic Landauer plateaux appear as expected.
As the inelastic scattering rate $1/\tau_{\rm in}$
progressively exceeds the elastic rate $1/\tau_{\rm el}$
(always keyed to the operational length of the structure), it is
also evident that there is a progressive loss of ideality.
Nonetheless the Landauer steps survive robustly,
albeit at a reduced height commensurate with the
degree of inelasticity.

To date, non-ideal behaviour in $G$ has been viewed
practically as an experimental nuisance, detracting
from the aim of detecting the perfect Landauer prediction
in ballistic wires
\cite{vanw,depic}.
On the contrary, we suggest that the observed deviations
from the ideal, for actual mesoscopic samples, carry valuable
information on non-equilibrium transport effects. That
these departures can, and should be, be probed systematically
follows from the logic of the microscopic analysis
presented above, supplemented with further detailed modelling
of the collision terms entering into Eqs. (\ref{eq1}) and (\ref{eq9}).

Our results, obtained from the standard Kubo-Greenwood theory and,
equally well, from the solution of a standard kinetic equation,
show how the Landauer conductance formula arises directly from
a fine-scale interplay of elastic and inelastic processes
in one-dimensional ballistic conductors. Such a derivation
automatically respects charge conservation and the
fluctuation-dissipation theorem. The latter is a natural outcome
of the analysis, not an additional hypothesis to be imposed ad hoc.

We have shown that the Landauer theory's traditional
phenomenological assumptions are not required for the validity
of the formula itself, provided the essential physics
of {\em resistive energy dissipation} is respected. Once the
inelastic processes responsible for dissipation are properly
included, the scope and value of the Landauer conductance
formula extend well beyond Landauer's original conception.
A minimal set of assumptions, as befits any microscopically
based approach, is not only enough to recover the full
Landauer formula; it also reveals considerably more information.

Finally, one conclusion stands out. In a mesoscopic ballistic
conductor open to its electrical environment, the close
interaction between dissipative and elastic scattering governs
the behaviour of the conductance. It does so uniquely. Neither of
the collision modes, acting alone, can sustain the physics
of mesoscopic transport. A theory of transport must allow all
such processes to act in concert, as they do in nature.

\end{document}